\documentclass[aip,apl,numerical]{revtex4-1}

\usepackage{graphicx}%
\usepackage{bm}%
\usepackage{amssymb}
\usepackage[amssymb]{SIunits}
\usepackage[version=3]{mhchem}

\def\Journal #1,#2,#3,#4#5#6#7{#1 {\bf #2}, #3 (#4#5#6#7)}

\newcommand{\CNRSAddress}{CEA/CNRS/UJF team “Nanophysics and semiconductors”,
Institute N$\acute{e}$el/CNRS-UJF, Grenoble, France}
\newcommand{\kistddress}{Center for Opto-Electronic Convergence Systems, Korea Institute of Science and Technology, Seoul 136-791, Korea}
\newcommand{\moonad}{Department of Physics, Tohoku University, Sendai 980-8578, Japan}
\newcommand{\leead}{Department of Emerging Materials Science, Daegu Gyeongbuk Institute of Science and Technology, Daegu 711-893, Korea}
\newcommand{\kimad}{Department of Physics, Yeungnam University, Gyeonsan 712-749, Korea}

\begin{document}
\title{The Effects of Post-Thermal Annealing on the Emission Spectra of GaAs/AlGaAs Quantum Dots grown by Droplet Epitaxy}

\author{Pilkyung Moon}
\email[Electronic mail: ]{pilkyung.moon@cmpt.phys.tohoku.ac.jp}
\affiliation{\moonad}
\author{J. D. Lee}
\affiliation{\leead}
\author{S. K. Ha }
\affiliation{\kistddress}
\author{E. H. Lee}
\affiliation{\kistddress}
\author{W. J. Choi}
\affiliation{\kistddress}
\author{J. D. Song}
\affiliation{\kistddress}
 
\author{J. S. Kim}
\affiliation{\kimad}
\author{L. S. Dang}
\affiliation{\CNRSAddress}

\date{\today{}}

\begin{abstract}

We fabricated GaAs/AlGaAs quantum dots by droplet epitaxy method, and obtained the geometries of the dots from scanning transmission electron microscopy data. Post-thermal annealing is essential for the optical activation of quantum dots grown by droplet epitaxy. We investigated the emission energy shifts of the dots and underlying superlattice by post-thermal annealing with photoluminescence and cathodoluminescence measurements, and specified the emissions from the dots by selectively etching the structure down to a lower layer of quantum dots. We studied the influences of the degree of annealing on the optical properties of the dots from the peak shifts of the superlattice, which has the same composition as the dots, since the superlattice has uniform and well-defined geometry. Theoretical analysis provided the diffusion length dependence of the peak shifts of the emission spectra.

\end{abstract}


\maketitle

Droplet epitaxy (DE)~\cite{Koguchi_et_al_1991a} is a novel growth method based on molecular beam epitaxy, which allows for the fabrication of a large variety of nanostructures with different geometries ranging from quantum dots (QDs),~\cite{Keizer_et_at_2010a} quantum dot molecules,~\cite{Yamagiwa_et_al_2006a,Romraeke_et_al_2008a} to single ring, and concentric rings.~\cite{Mano_et_al_2005a} Unlike the Stranski-Krastanov method, DE enables the formation of QDs even with materials with small lattice mismatch.~\cite{Mantovani_et_al_2004a} However, for QDs grown by DE, post-thermal annealing is essential to optically activate the transitions, due to the relatively large numbers of defects in as-grown samples. Several papers have reported significant changes in the optical spectra of QDs by post-thermal annealing.~\cite{Park_et_al_2007a,Zhuang_et_al_2000a,Choi_et_al_2011a,Sanguinetti_et_al_2002a}
However, only few articles have reported the structural changes of DE QDs and the diffusion length, which directly affect the QD peak shift during post-thermal annealing, from the QD peak itself.~\cite{Sanguinetti_et_al_2002a} Thus, more in-depth study is necessary for real application of DE QDs.

In this paper, we experimentally and theoretically investigate the effects of post-thermal annealing on the emission energies of GaAs/AlGaAs QDs grown by DE. We specify the emission peaks from QDs by selectively etching the sample and comparing the cathodoluminescence spectra in the regions with and without QDs. In addition, we show that the diffusion length of GaAs due to the annealing process can be obtained from the emission peak shift of the underlying GaAs/AlGaAs superlattice (SL) with high precision. We apply the diffusion to the QD model obtained by scanning transmission electron microscopy (STEM), and show the good agreement between the theoretical results and experimental measurements.

\begin{figure*}
\begin{center}
\leavevmode\includegraphics[width=0.9\hsize]{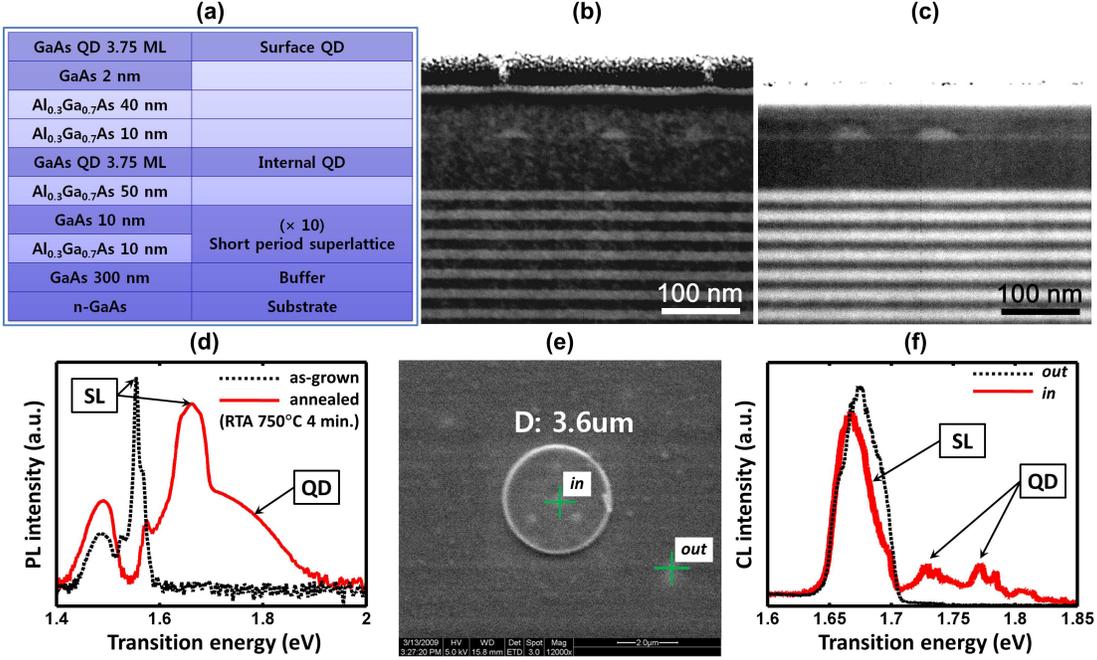}
\end{center}
\caption{ (Color online) (a) Structure of our GaAs/$\rm Al_{0.3}Ga_{0.7}As$ DE QDs and short period SL. Scanning transmission electron microscopy data of (b) as-grown and (c) annealed dots. (d) Photoluminescence spectra of the sample taken at 15.8\,K before (black dashed line) and after (red solid line) the annealing. (e) Field emission scanning electron microscopy data of the mesa-etched structure. (f) 5.3\,K-cathodoluminescence spectra inside (red solid line) and outside (black dashed line) the mesa. Annealed sample was used for (e) and (f).}
\label{f1}
\end{figure*}

Figure \ref{f1}(a) shows the structure of our GaAs/$\rm Al_{0.3}Ga_{0.7}As$ DE QDs and short period SL. We fabricated the samples on n-typed GaAs (001) wafers. After the growth of a 300\,nm-thick GaAs buffer, 10 alternating repetitions of a 10\,nm-thick $\rm Al_{0.3}Ga_{0.7}As$ and a 10\,nm-thick GaAs short period SL, and a 50\,nm-thick $\rm Al_{0.3}Ga_{0.7}As$ layer were grown successively. Then, GaAs QDs were grown by droplet epitaxy. Details of the growth sequence are described elsewhere.~\cite{Ha_et_al_2011a} After the growth of the dots, a 10\,nm-thick $\rm Al_{0.3}Ga_{0.7}As$ layer was grown by a migration enhanced method, and an additional 40\,nm-thick $\rm Al_{0.3}Ga_{0.7}As$ layer and a 2\,nm-thick GaAs layer were grown successively. We annealed the sample at $750\,^{\circ}\mathrm{C}$ for 240\,s in an $\rm N_{2}$ atmosphere to improve the optical properties of the dots. Figures \ref{f1}(b) and \ref{f1}(c) show the cross-sectional STEM data of the as-grown and annealed samples, respectively. The striped patterns in the lower half of the figures represent the GaAs/AlGaAs SL, while the lens-shaped objects in the upper half show the QDs. We obtained the average height 9.05\,nm and width 35.6\,nm of the as-grown dots from STEM data, and the in-plane density of dots $\sim$4$\times$10$^{9}$/cm$^{2}$ from atomic force microscopy data. The dashed (black) and solid (red) lines in Fig.\ \ref{f1}(d) shows the photoluminescence spectra of the sample before and after the annealing, respectively. 
Since the QDs grown by DE have poor luminescence properties in the absence of post-thermal annealing,~\cite{Park_et_al_2007a,Zhuang_et_al_2000a,Choi_et_al_2011a,Sanguinetti_et_al_2002a} the two closely spaced sharp peaks near 1.56\,eV of the as-grown sample originate from the GaAs/AlGaAs SL.
The post-thermal annealing shifts the SL peaks to higher energies at around 1.65\,eV, and broadens the linewidth of the peaks. In addition, the photoluminescence spectra show that the post-thermal annealing activates the transitions at 1.7$\sim$1.8\,eV, which are apparently distinguishable from the SL peaks. To investigate the physical origin of the transitions, we etched the annealed sample to the lower layer of the dots, while leaving a mesa structure with a diameter of \unit{3.6}{\micro\meter}. Figure 1(e) shows the field emission scanning electron microscopy data of the mesa structure (denoted as \textbf{\textit{in}}), and etched region (denoted as \textbf{\textit{out}}). In Fig.\ \ref{f1}(f), we plot the cathodoluminescence spectra from the mesa (red solid line; \textbf{\textit{in}}), and the etched region (black dashed line; \textbf{\textit{out}}), respectively. The peaks at 1.67\,eV are observed from both the mesa and etched region, while the transitions over 1.7\,eV are observed only at the mesa. Since QDs exist only in the mesa, we concluded that the transitions over 1.7\,eV in Figs.\ \ref{f1}(d) and \ref{f1}(f) originate from the GaAs QDs.

\begin{figure}
\begin{center}
\leavevmode\includegraphics[width=0.9\hsize]{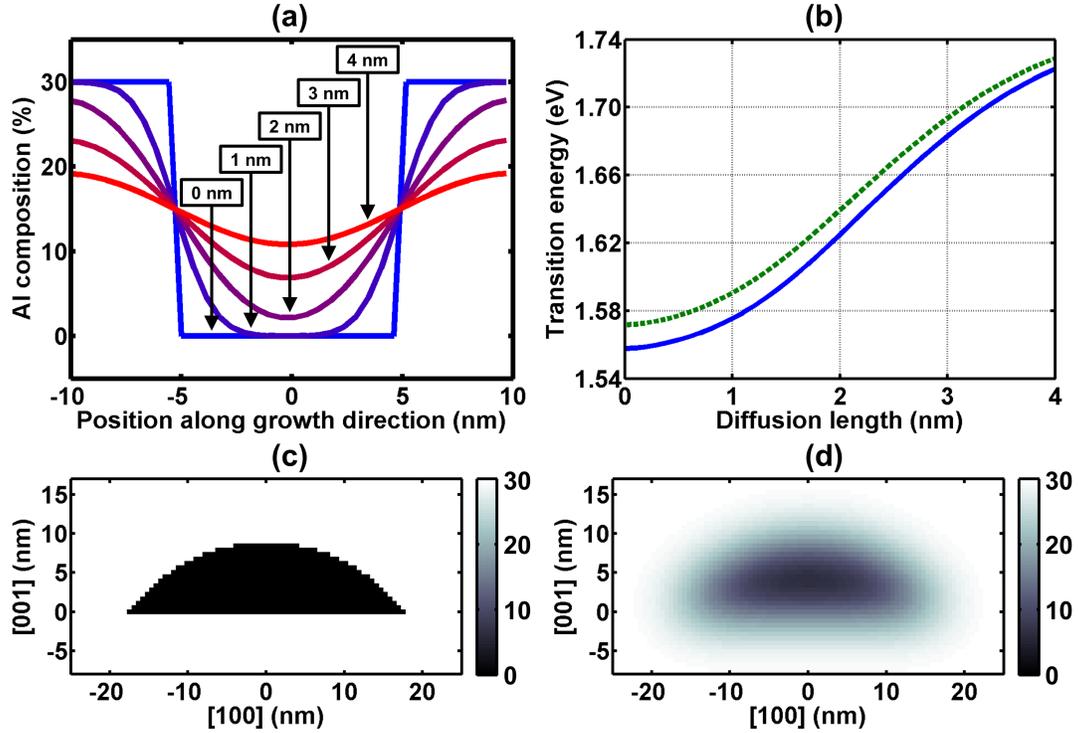}
\end{center}
\caption{ (a) Al composition profiles of SL along the growth direction for various diffusion lengths. (b) The first (blue solid line) and the second (green dashed line) transition energies of SL in varying diffusion lengths. Al composition of QD (a) before and (b) after the annealing.}
\label{f2}
\end{figure}

\begin{figure*}
\begin{center}
\leavevmode\includegraphics[width=0.9\hsize]{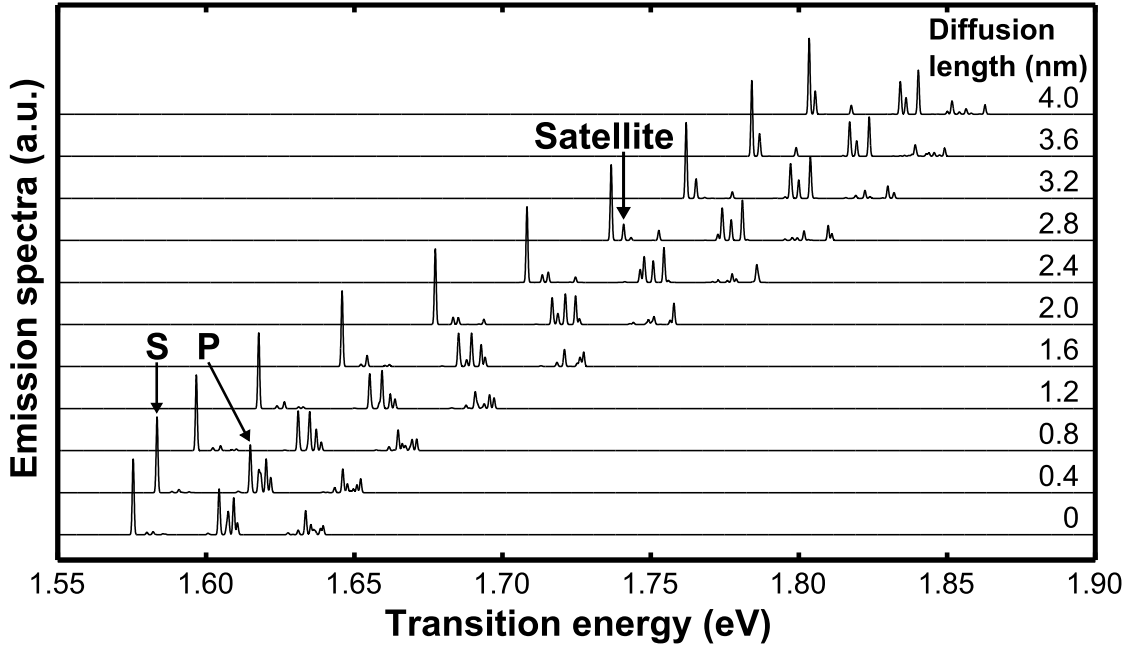}
\end{center}
\caption{Calculated emission spectra of DE QDs in varying diffusion lengths.}
\label{f3}
\end{figure*}

We performed numerical simulations to investigate the emission energy shifts of the DE QDs by the annealing process. We calculated the electronic structures and exciton spectra of the dots, by using an eight–band k$\cdot$p method, and a configuration interaction method, respectively.~\cite{Moon_et_al_2009a} The effects of post-thermal annealing are introduced by the diffusion length of GaAs. Since both the SL and dots are composed of the same material (GaAs), with the same barrier material ($\rm Al_{0.3}Ga_{0.7}As$), we can obtain the diffusion length of the dots from the peak shifts of the SL.
This method has an advantage that the diffusion length can be obtained with high accuracy, since SL can be fabricated with much higher geometric precision than QDs.
We plot the Al composition profiles of SL along the growth direction for five different diffusion lengths of 0, 1, 2, 3, and 4\,nm in Fig.\ \ref{f2}(a). The solid (blue) and dashed (green) lines in Fig.\ \ref{f2}(b) show the two lowest emission energies of the SL in varying diffusion lengths, which correspond to the transitions from the electron ground state to the first heavy-hole and light-hole states, respectively. The theoretically estimated SL emission energies of 1.558 and 1.572\,eV in the absence of diffusion [diffusion length = 0\,nm in Fig.\ \ref{f2}(b)] agree well with the experimentally observed values of 1.55 and 1.57\,eV [the two closely spaced sharp peaks in Fig.\ \ref{f1}(d)].
We calculated the diffusion length of 2.4\,nm from the experimentally observed SL peak after the annealing [Fig.\ \ref{f1}(d)] and the relation between the emission energies of SL and diffusion lengths [Fig.\ \ref{f2}(b)].
We applied the obtained diffusion length to a QD model estimated from the STEM data [Fig.\ \ref{f1}(b)]. Figures \ref{f2}(c) and \ref{f2}(d) show the Al composition profiles of the AlGaAs dot before, and after the annealing, respectively. Due to the lens-shaped geometry of the dot, the composition grading below the dot is much larger than that above the dot.~\cite{Moon_et_al_2011a} We plot the calculated emission spectra of the dot with varying diffusion lengths in Fig.\ \ref{f3}. In spite of the lack of biaxial strain between GaAs and AlGaAs, the heavy-hole and light-hole states still exhibit small but finite splitting, due to the difference between the effective masses of the holes.
In addition, the energy spacing between the electron states is much larger than that between the hole states since holes are much heavier than electrons. Thus, the peaks from QDs for each diffusion length are classified into several groups; s-channel (marked as {\bf S}), p-channel (marked as {\bf P}), and so forth. The post-thermal annealing with a diffusion length 2.4\,nm shifts the {\bf S} and {\bf P} emission peaks from 1.575 and 1.607\,eV (diffusion length = 0\,nm) to 1.708 and 1.751\,eV. Thus, the theoretical result is in good agreement with the fact that the peaks above 1.7\,eV in Figs.\ \ref{f1}(d) and \ref{f1}(f) originate from the annealed DE QDs.
As can be seen from Fig.\ \ref{f3}, the amount of the emission energy shift does not linearly vary with diffusion length. The amount of the shift exhibits a maximum at around the diffusion length 2.4\,nm, and becomes smaller as the diffusion length deviates from 2.4\,nm.
Such a non-linear shift of the QD peak is consistent with the peak shift of SL in Fig.\ \ref{f2}(b), and can be explained by the composition profiles of SL [Fig.\ \ref{f2}(a)].
The composition grading mainly occurs at the boundary for small diffusion (diffusion length $\leq$ 1.0\,nm), while the whole area of the layer (and dot) becomes graded as the diffusion length increases. Thus, the amount of the peak shift increases with the diffusion length. However, as the diffusion length increases over 3.0\,nm, the quantum confinement energy considerably reduces, thus the peak shift gradually decreases.
The energy difference between the ground emission peak {\bf S} and the first excited emission peak {\bf P} increases as the diffusion length increases, up to the diffusion length of 2.4\,nm, then decreases, as the diffusion length increases. On the contrary, the energy difference between the ground peak {\bf S} and its satellite peak (marked as {\bf Satellite}) monotonically decreases, as the diffusion length increases. In addition, as the diffusion length increases, the emission strength of the satellite peak grows, and the number of p-channels reduces from four to three.

In conclusion, we fabricated GaAs/AlGaAs QDs by DE, and investigated the effects of post-thermal annealing on the emission energies. The analysis of photoluminescence and cathodoluminescence data with mesa etched structure reveals that the post-thermal annealing process considerably shifts the emission peaks of the dots. We calculated the diffusion length of our GaAs/$\rm Al_{0.3}Ga_{0.7}As$ system from the photoluminescence peak shifts of the underlying SL. By applying the obtained diffusion length 2.4\,nm to the QD model estimated from the STEM data, we confirmed that the peaks above 1.7\,eV in the annealed sample originate from the diffused GaAs DE QDs. Theoretical analysis provided the diffusion length dependence of the emission energies of QDs and SL. We show that the amount of the peak shift exhibits a maximum at around a specific diffusion length, and reduces as the diffusion length deviates from that value, due to the competition between the blueshift by composition grading and the redshift by quantum confinement. As a result, more careful control of annealing conditions are necessary for the post-thermal annealing in this regime.

This work was partially supported by Special Coordination Funds for Promoting Science and Technology from MEXT, Japan. P.M. would like to acknowledge the support from the KISTI Supercomputing Center through the strategic support program for the supercomputing application research (KSC–2009–S01–0009). The authors in KIST acknowledge the support from the KIST institutional program, the Converging Research Center Program through the MEST (2012K001280), DAPA,  ADD and GRL program.


\begin{thebibliography}{99}

%\bibitem{barve2010systematic} A. Barve, T. Rotter, Y. Sharma, S. Lee, S. Noh, and S. Krishna, Appl. Phys. Lett. {\bf 97}, 061105 (2010).




\bibitem{Koguchi_et_al_1991a} N. Koguchi, S. Takahashi, and T. Chikyow, \Journal J. Cryst. Growth,111,688,1991.

\bibitem{Keizer_et_at_2010a} J. G. Keizer, J. Bocquel, P. M. Koenraad, T. Mano, T. Noda, and K. Sakoda, \Journal Appl. Phys. Lett.\null,96,062101,2010.

\bibitem{Yamagiwa_et_al_2006a} M. Yamagiwa, T. Mano, T. Tateno, K. Sakoda, G. Kido, N. Koguchi, and F. Minami, \Journal Appl. Phys. Lett.\null,89,113115,2006.

\bibitem{Romraeke_et_al_2008a} R. Pomraeke, C. Lienau, Y. I. Mazur, Z. M. Wang, B. Liang, G. G. Tarasov, and G. J. Salamo, \Journal Phys. Rev. B,77,075314,2008.

\bibitem{Mano_et_al_2005a} T. Mano, T. Kuroda, S. Sanguinetti, T. Ochiai, T. Tateno, J. Kim, T. Noda, M. Kawabe, K. Sakoda, G. Kido, and N. Koguchi, \Journal Nano Letters,5,425,2005.

\bibitem{Mantovani_et_al_2004a} V. Mantovani, S. Sanguinetti, M. Guzzi, E. Grilli, M. Gurioli, K. Watanabe, and N. Koguchi, \Journal J. Appl. Phys.\null,96,4416,2004.

\bibitem{Park_et_al_2007a} H. J. Park, J. H. Kim, J. J. Yoon, J. S. Son, D. Y. Lee, H. H. Ryu, M. H. Jeon, and J. Y. Leem, \Journal J. Cryst. Growth,300,319,2007.

\bibitem{Zhuang_et_al_2000a} Q. D. Zhuang, J. M. Li, X. X. Wang, Y. P. Zeng, Y. T. Wang, B. Q. Wang, L. Pan, J. Wu, M. Y. Kong, and L. Y. Lim, \Journal J. Cryst. Growth,208,791,2000.

\bibitem{Choi_et_al_2011a} Hyun Young Choi, Min Young Cho, Min Su Kim, Jae-Young Leem, Dong-Yul Lee, Jin Soo Kim, and Jong Su Kim, \Journal J. Korean Phys. Soc.\null,,58,1324,2011.

\bibitem{Sanguinetti_et_al_2002a} S. Sanguinetti, K. Watanabe, T. Kuroda, F. Minami, Y. Gotoh, and N. Koguchi, \Journal J. Cryst. Growth,242,321,2002.

\bibitem{Ha_et_al_2011a} Seung-Kyu Ha, Jin Dong Song, Su Youn Kim, Jung Il Lee, Samir Bounouar, Le Si Dang, and Jong Su Kim, \Journal J. Korean Phys. Soc.\null,58,1330,2011.

\bibitem{Moon_et_al_2009a} P. Moon, E. Yoon, W. Sheng, and J.-P. Leburton, \Journal Phys. Rev. B,79,125325,2009.

\bibitem{Moon_et_al_2011a} P. Moon, S.-K. Ha, J. D. Song, J. Y. Lim, S. Bounouar, F. Donatini, L. S. Dang, J. P. Poizat, J. S. Kim, W. J. Choi, I. K. Han, and J. I. Lee, \Journal AIP Conf. Proc.\null,1399,425,2011.

\end{thebibliography}
\end{document}